\def\cZ{Z}
\def\bardelta{ \delta '}
\def\ins#1{}

\def\hbeta{{ \beta }}

\def\comment#1{}

\def\cm#1{}

\def\>{\rangle}
\def\<{\langle}
\def\comment#1{}

\documentclass[pra,aps]{article}
\usepackage{graphics}
\begin{document}
\title{Perturbatively Defined Effective Classical Potential\\ in Curved Space
%Perturbation Expansion of Path Integrals
%%in Curved Space-Zero Modes for Periodic Boundary Conditions
}
\author{H.~Kleinert and  A.~Chervyakov\\
Institut f\"ur Theoretische Physik\\ Freie Universit\"at Berlin,\\
Arnimallee 14, 14195 Berlin, Germany
\\~\\
}
%\\
%\pacs{nn.mm.xx}{First pacs description}
%\pacs{nn.mm.xx}{Second pacs description}
%\pacs{nn.mm.xx}{Third pacs description}
%\begin{document}
\maketitle

\begin{abstract}
The partition function of a quantum statistical system
in flat space can always be written as an integral over
a classical Boltzmann factor
$\exp[ -\beta V^{\rm eff\,cl}({\bf x}_0)]$,
where $V^{\rm eff\,cl}({\bf x}_0)$
is the so-called {\em effective classical potential\/}
containing the effects of all quantum fluctuations.
The variable of integration is
the temporal path average
${\bf x}_0\equiv \beta ^{-1}\int _0^ \beta d\tau \,{\bf x}(\tau )$.
We show how to generalize  this concept
to paths $q^\mu(\tau )$ in curved space with metric $g_{\mu \nu }(q)$,
and calculate perturbatively the
high-temperature expansion of $V^{\rm eff\,cl}(q_0)$.
The requirement of
independence under coordinate transformations
$q^\mu(\tau )\rightarrow q'^\mu(\tau )$
introduces subtleties
in  the definition and treatment of the path average $q_0^\mu$,
and covariance is achieved only
with the help of a suitable Faddeev-Popov procedure.
\end{abstract}
\section{Introduction}
Path integrals for particles in  curved space
are defined unambiguously
as nonholonomic images of flat-space path integrals
(the procedure following the
so-called {\em nonholonomic mapping principle\/}) \cite{ins1}.
The resulting time evolution amplitudes satisfy
automatically the correct Schr\"odinger equation
{\em without\/} an extra $R$-term \cite{insHR}, and they are
invariant under arbitrary coordinate transformations.
%which do not change the geometry.
For perturbatively defined path integrals,
a similar implementation of the
nonholonomic mapping principle
in curved space has so far not been found.
There the absence of an extra $R$-term must be
inferred from the above time-sliced
theory. Even coordinate invariance
was a problem for a long time, resolved
only recently by our treatment via good-old dimensional
regularization \cite{ins2}. Moreover, we were able to
find well-defined calculation rules
for dealing with products of distributions,
which enable us now to perform perturbation expansions
{\em without\/} a tedious extension of the dimension of the time axis.
Due to the local nature of these rules, they can be applied
to Feynman integrals with infinite as well as
finite propagation times \cite{ins3,ins4}.

One important aspect of perturbatively defined path integrals
with a finite propagation time has, however, remained  puzzling.
The results have so far been found correctly only
if calculations are done with propagators
satisfying Dirichlet boundary conditions.
Attempts with periodic paths
have led to noncovariant results \cite{ins5}.

Some years ago it has been pointed out by Feynman
and one of the authors (H.K.) that in flat space,
the temporal average
${\bf x}_0\equiv \bar {\bf x}(\tau ) =
\beta ^{-1}\int_0^\beta d\tau \,{\bf x}(\tau )$
of periodic paths plays a special role in isolating
the classical fluctuations in a path integral over periodic paths \cite{insx0}.

An ordinary integral over ${\bf x}_0$
which has the form of a classical partition function
can produce the full quantum statistical result, if it is performed over a
Boltzmann factor containing the so-called
{\em effective classical potential\/} $V^{\rm eff\,cl}({\bf x}_0)$.
If a similar quantity is calculated in curved space
keeping the temporal average $q_0 \equiv  \bar q(\tau )
\equiv \beta ^{-1}\int_0^\beta d\tau  q(\tau )$
fixed, the two-loop perturbative result
for $V^{\rm eff\,cl}(q_0)$ turned out to
deviate from the covariant one by a noncovariant
total derivative \cite{ins5}, in contrast to the covariant result
obtained with Dirichlet boundary conditions.
For this reason, perturbatively defined path integrals
with periodic boundary conditions
in curved space have been of limited use
in the presently popular first-quantized worldline approach
to quantum field theory
(also called the string-inspired approach reviewed in Ref.~\cite{ins6}).
In particular,
is has so far been impossible to
calculate
with periodic boundary conditions
interesting quantities such as
curved-space effective actions,
gravitational anomalies, and
index densities
\cite{ins7},
 all results
having been reproduced
with Dirichlet boundary conditions \cite{ins6}.

The purpose of this paper is to improve the situation
by developing a manifestly covariant
integration procedure for periodic paths.
It is an adaption of similar procedures used before in the effective
action formalism of two-dimensional sigma-models \cite{ins8,ins9}.
Covariance is achieved by expanding the fluctuations
in the neighborhood of any given point in powers
of geodesic coordinates, and by a covariant definition of a path average
different from the naive temporal average.
As a result, we shall find the same locally covariant perturbation expansion
of the effective classical potential as
in earlier work with Dirichlet boundary conditions \cite{ins3}.

An important role in the development is played by
the Faddeev-Popov method, which produces a
Jacobian and an associated new effective interaction
necessary to guarantee covariance.

%%%%%%%%%%%%%%%%%%%%%%%%
\section{Partition function}
%%%
Consider a quantum particle moving in a compact Riemannian space
with metric $g_{\mu \nu } (q)$ and coordinates
$q^\mu (\tau ),~\mu=1, \dots, D$.
The partition function can be written as an
integral over the {\em partition function density\/}  $z(q)$:
\begin{eqnarray}
{\cZ} & = & \int d^D q  \sqrt{g(q)} \,z(q),
\label{ins2.1}\end{eqnarray}
where $g  = \det g_{\mu \nu }$.
The partition function density
is equal to the diagonal time evolution amplitude
$\langle q^\mu  \beta  \vert q^\mu 0 \rangle $,
and has the path integral representation
\begin{eqnarray}
z(q)= \langle q^\mu  \beta  \vert q^\mu 0 \rangle & = &
\int^{q^\mu (\beta) = q^\mu}_{ q^\mu (0) = q^\mu}
{\cal D}^D q (\tau )  \sqrt{g (q(\tau ))} e^{{\cal A}_{\rm e} [q]} ,
\label{ins2.2}\end{eqnarray}
with the euclidean action
\begin{eqnarray}
 {\cal A}_{\rm e} [q] = \int_{0}^{ \beta } d\tau  \frac{1}{2}
  g_{\mu \nu } (q) \dot q^\mu (\tau ) \dot q^ \nu  (\tau ).
\label{ins2.3}\end{eqnarray}
The invariant measure represents formally the product
\begin{eqnarray}
{\cal D}^D  q  \sqrt{g (q)} \equiv   \prod_{\mu,\tau }
   \left[ d q^\mu (\tau )  \sqrt{g (q(\tau ) )}\right].
\label{ins2.4}\end{eqnarray}
In our notation, a single symbol $ \sqrt{g (q(\tau ) )}$
in the measure on the left-hand side
symbolizes a factor
$ \sqrt{g (q(\tau ) )}$ for {\em each\/} time point.

For small inverse temperature $\beta$, the path integral
(\ref{ins2.2}) can be calculated perturbatively using Green function
with Dirichlet boundary conditions, leading to
a manifestly covariant high-temperature expansion,
whose initial terms are \cite{ins3,ins4,ins1}
\begin{eqnarray}
\langle q^\mu  \beta  \vert q^\mu 0\rangle =
\frac{1}{ \sqrt{2 \pi  \beta }^D } \left[ 1 - \frac{1}{24} R
 (q)  \beta  + \dots \right] .
\label{ins2.5}\end{eqnarray}
This differs from the well-known quantum-mechanical DeWitt-Seeley expansion
of the exponential of the Laplace-Beltrami operator
$\Delta={ \sqrt{g} }^{-1}\partial _\mu \sqrt{g}g^{\mu \nu}\partial_\nu$,
\begin{eqnarray}\!\!\!\!\!\!\!
\langle q^\mu|e^{ \beta  \Delta /2}|q^\mu\rangle
 =
\frac{1}{ \sqrt{2 \pi  \beta }^D } \left[ 1 + \frac{1}{12} R
 (q)  \beta  + \dots \right]
\label{ins2.5S}\end{eqnarray}
by a term $R (q) \beta /8$ in the brackets.
Since the correct path integral defined
by the nonholonomic mapping principle in Ref.~\cite{ins1}
agrees with (\ref{ins2.5S}), the invariant volume element (\ref{ins2.4})
of path integration must be  corrected
by a factor $\exp[\int^{ \beta }_{0}d\tau  R (q)/8]$.
%(see the detailed discussion in \cite{ins1}).
%It is, however, beyond the scope of this note.

For consistency of the perturbative approach,
we must of course be able to calculate
the same partition function (\ref{ins2.1})
with the same result (\ref{ins2.5})
by performing a functional integral over all periodic paths
\begin{eqnarray}
 {\cZ}^{\rm P} & = &\oint
%_{ q^\mu ( \beta ) = q^\mu (0)}
{\cal D}^D q  \sqrt{g (q)}
e^{-{\cal A}_{\rm e} [q] }\,,
\label{ins2.6}\end{eqnarray}
where the symbol $\oint$ indicates the
periodicity of the paths.
A method for doing this  will now
be developed.
%%%%%%
\section{Covariant fluctuation expansion}
For small $\beta$, the path integral (\ref{ins2.6})
is dominated by the constant paths $ q^\mu (\tau ) = q^\mu_0$
which
are solutions of the classical equations of motion
whose classical action vanishes.
To derive the high-temperature expansion of the path integral (\ref{ins2.6}),
we parametrize the small fluctuations around $q_0^\mu $
covariantly by geodesic coordinates $\xi^\mu(\tau )$.
This is done with a {\em nonlinear\/} decomposition
\begin{eqnarray}
  q^\mu (\tau ) = q^\mu_0 + \eta^\mu (q_0, \xi),
\label{ins3.1}\end{eqnarray}
where $ \eta ^\mu(q_0, \xi)=0$
for $\xi^\mu=0$.
The geodesic coordinates $\xi^\mu (\tau )$ are the tangent vectors
at $q^\mu_0$ to the geodesic connecting the points
$q^\mu_0$ and $q^\mu_0 + \eta^\mu$.
The functions $\eta^\mu (q_0, \xi)$ have the expansion
\begin{eqnarray}
\eta ^\mu (q_0, \xi)= \xi^\mu - \frac{1}{2}  \Gamma_{(\sigma \tau) }{}^\mu
(q_0) \xi^\sigma \xi^\tau
- \frac{1}{6} \Gamma _{(\sigma \tau \kappa)}{}^\mu
   (q_0) \xi^ \sigma \xi^\tau  \xi^\kappa - \dots,
\label{ins3.2}\end{eqnarray}
where $\Gamma _{ \sigma \tau }{}^\mu (q_0)$  with two lower indices
is the usual
Christoffel symbol, while  the other
coefficients $\Gamma _{ \sigma \tau \dots \kappa }{}^\mu (q_0)$
are its successive covariant derivatives with respect to lower indices only,
\begin{eqnarray}
 \Gamma _{ \sigma \tau  \kappa }{}^\mu
 (q_0) = \nabla_ \kappa
  \Gamma _{ \sigma \tau }{}^\mu
 = \partial_ \kappa  \Gamma _{ \sigma \tau }{}^\mu
 - 2  \Gamma_{ \kappa  \sigma } {}^\nu
 \Gamma _{ \nu \tau }{}^\mu
, \dots
\label{ins3.3}\end{eqnarray}
The parentheses around the subscripts in
(\ref{ins3.2}) indicate symmetrization
with respect to  all possible cyclic permutations.

If the initial coordinates
$q^\mu$ are themselves geodesic at $q_0^\mu$,
all coefficients $ \Gamma _{( \sigma \tau \dots \kappa )}{}^\mu (q_0)$
in Eq.~(\ref{ins3.2}) are zero, so that
$\eta^ \mu  (\tau ) = \xi^\mu(\tau )$, and the decomposition
(\ref{ins3.1}) is linear. In this case, the derivatives of the
Christoffel symbols can be expressed directly
in terms of the curvature tensor:
\begin{eqnarray}
\partial_ \kappa   \Gamma ^{\mu}_{\tau  \kappa }
(q_0) = - \frac{1}{3}  \left[ R_{\tau  \kappa  \sigma }{}^\mu (q_0)
+ R_{ \sigma  \kappa \tau }{}^\mu (q_0) \right],
~~~~{\rm for~normal~coordinates}.
\label{ins3.4}\end{eqnarray}
In arbitrary coordinates, however, $\eta^\mu (\tau )$
does not transform like a vector under coordinate
transformations, and we must use the nonlinear decomposition (\ref{ins3.1}).

We now transform the path integral (\ref{ins2.6})
to the new coordinates $\xi^\mu (\tau )$ using
Eqs.~(\ref{ins3.1})--(\ref{ins3.3}). %
%\footnote{All these general constructions are
%the same as in the covariant effective action
%formalism\cite{ins8}.}
The perturbation
expansion for the transformed path integral over $\xi^\mu(\tau )$
is constructed for any chosen $  q_0^\mu$
by expanding the action (\ref{ins2.3}) and the measure (\ref{ins2.4})
in powers of small linear fluctuations $\xi  ^\mu(\tau )$.
The expansion starts out like
\begin{eqnarray}
{\cal A}_{\rm e} [q]
 & = &
\int_0^ \beta
 d\tau \frac{1}{2}
 \left[ g_{\mu \nu } (q_0) + \partial_ \sigma g_{\mu \nu }
 (q_0) \eta^ \sigma  +
 \frac{1}{2} \partial_ \sigma  \partial_\tau
 g_{\mu \nu } (q_0) \eta^ \sigma  \eta ^\tau  + \dots \right]
 \dot \eta^\mu \dot \eta^\nu  \nonumber \\
  &  = &
\int_0^ \beta
 d\tau\frac{1}{2} \left[ g_{\mu \nu } (q_0) + \frac{1}{3} R_{\mu
\lambda_1 \nu \lambda_2} (q_0) \xi^{ \lambda _1} \xi^{ \lambda _2}
   + \dots \right]  \dot \xi^\mu \dot \xi^ \nu .
\label{ins3.5}\end{eqnarray}
The leading small-$\beta$ behavior of the path
integral (\ref{ins2.6}) is given by the quadratic term in $\xi^\mu(\tau )$,
which we write after a partial integration as
\begin{eqnarray}
 {\cal A}_{\rm e}^{(0)} [q_0 , \xi] = \int^{ \beta }_{0} d\tau  \frac{1}{2}
   \xi^\mu(\tau ) [-g_{\mu \nu } (q_0)d_\tau^2 ] \xi^ \nu(\tau ).
\label{ins3.6}\end{eqnarray}
The next term in $\beta$ is caused by the interaction
of fourth order in the fluctuations:
\begin{eqnarray}
 {\cal A}_{\rm e}^{{\rm int},4} [q_0, \xi] = \int^{ \beta }_{0} d\tau
\frac{1}{6} R_{\mu \lambda _1  \nu  \lambda _2} (q_0) \xi^{ \lambda _1}(\tau)
   \xi^{ \lambda _2}(\tau ) \dot \xi^\mu (\tau )\dot \xi^ \nu(\tau ) .
\label{ins3.7}\end{eqnarray}
A further contribution comes from the invariant measure (\ref{ins2.4}).
This is transformed to the coordinates $\eta^\mu(\tau )$ as
\begin{eqnarray}
 \!\!\!\!\!\!\!\!
\!\!\!\!\!\!\!\!
{\cal D}^D  q  \sqrt{g (q)} =
\prod_{\mu,\tau } \left[ dq^\mu (\tau )  \sqrt{g(q(\tau ))} \right]
 = \sqrt{g(q_0)}^N
 \prod _{\mu,\tau } \left[ d\eta^\mu (\tau )
 \frac{\sqrt{g(q_0
 + \eta (\tau ))}}{\sqrt{g(q_0)}}\right]\,,
\label{ins3.8a}
\end{eqnarray}
and further to the geodesic coordinates $\xi^\mu(\tau )$ as
\begin{eqnarray}
{\cal D}^D q \sqrt{g(q)} =
\sqrt{g(q_0)}^N J(q_0, \xi) \prod_{\mu,\tau }
\left[ d\xi^\mu (\tau ) \frac{ \sqrt{g (q_0 + \eta (q_0,\xi))} }
   { \sqrt{g(q_0)}  }\right] ,
\label{ins3.8}\end{eqnarray}
where $J(q_0,\xi)$ is the Jacobian of the transformation (\ref{ins3.2}):
\begin{eqnarray}
  J(q_0, \xi) = \exp\left\{ \int^{ \beta }_{0} d\tau\, \delta (\tau ,\tau )
\,{\rm tr}\hspace{-1pt}\log
\left(\frac{\partial \eta^\mu}{\partial \xi^ \nu } \right)
\right\} .
\label{ins3.9}\end{eqnarray}
The trace of the logarithm in the exponent
has the small-$\xi^\mu$ expansion
\begin{eqnarray}
%&& \log \det \left(\frac{\partial\eta^\mu}{\partial \xi^ \nu } \right) =
\!\!\!\!\!\!\! {\rm tr}\log \left(\frac{\partial \eta^\mu}{\partial \xi^ \nu } \right)
 %\nonumber \\&&
 \! = \!-  \Gamma _{\mu  \sigma  }^\mu  \xi^ \sigma
 + \frac{1}{3} \left(\frac{1}{2}  \Gamma _{ \nu \tau }^\mu
     \Gamma ^ \nu _{\mu \sigma } +  \Gamma ^ \nu _{\tau  \sigma }
  \Gamma ^\mu_{ \nu \mu} - \partial_ \sigma   \Gamma ^\mu_{\mu\tau }
  - \frac{1}{2} \partial_\mu  \Gamma ^\mu_{ \sigma \tau } \right)
   \xi^ \sigma
   \xi^\tau   +\dots~.
\label{ins3.10}\end{eqnarray}
The exponent contains also an infinite quantity
\begin{eqnarray}
 N = \int^{ \beta }_{0} d\tau \,\delta (\tau ,\tau ) =  \beta \, \delta (0),
\label{ins3.11}\end{eqnarray}
which formally represents the total number of
points on the time axis and counts simultaneously
the number of eigenvalues of the
operator $-g_{\mu \nu } (q_0)d_\tau^2 $
in the space of  periodic functions $\xi^\mu(\tau )$.
By rewriting also the product on the right-hand side
of Eq.~(\ref{ins3.8}) as an exponential
\begin{eqnarray}
  \prod_{\tau } \frac{ \sqrt{g(q_0 + \eta (q_0,\xi))} }{ \sqrt{g(q_0)} }
  =\exp\left\{ \int^{ \beta }_{0}d\tau\,  \delta (\tau ,\tau )\frac{1}{2}
  \log \frac{g(q_0 + \eta (q_0,\xi))}{g(q_0)} \right\} ,
\label{ins3.12}\end{eqnarray}
and expanding
\begin{eqnarray}
 \!\!\!\!\!\!\!\!\!\!\!\!\!\!\!\!\!\!\!\!\!\!\!
\frac{1}{2} \log \frac{g(q_0 + \eta )}{g(q_0)}
= \Gamma ^\mu_{\mu \sigma } \eta^ \sigma\! +\! \frac{1}{2}
 \partial_ \sigma   \Gamma ^\mu_{\tau \mu} \eta^ \sigma \eta^\tau
+ \dots~\label{ins3.13a}, ~~~~
\end{eqnarray}
and this further into
 \begin{eqnarray}
\frac{1}{2} \log \frac{g(q_0 + \eta (q_0,\xi))}{g(q_0)}
 =\Gamma ^\mu_{\mu  \sigma } \xi^ \sigma \!
+\! \frac{1}{2}
\left(\partial_ \sigma   \Gamma _{\tau \mu}^{\mu}\! -\!\Gamma ^\mu_{ \nu \mu}
 \Gamma ^ \nu _{ \sigma \tau }\right) \xi^ \sigma \xi^\tau+\dots ,
\label{ins3.13}\end{eqnarray}
we may combine the expansions (\ref{ins3.12}),  (\ref{ins3.13})
with (\ref{ins3.9}), (\ref{ins3.10}), and obtain
\begin{eqnarray}
\!\!\!\!\!\!\!\!\!\prod_{\mu,\tau } \left[ dq^\mu (\tau )\sqrt{g(q(\tau ))}
\right]  =  \sqrt{g(q_0)} ^N \prod_{\mu,\tau } \left[
d\xi^\mu(\tau ) \right]\exp\left\{-{\cal A}_{\rm e}^{{\rm meas}}
[q_0,\xi] \right\} ,
\label{ins3.14}\end{eqnarray}
where
${\cal A}_{\rm e}^{{\rm meas}} [q_0, \xi]$ plays the role
of an interaction coming from the invariant measure.
Its expansion starts out like
\begin{eqnarray}
 {\cal A}_{\rm e}^{{\rm meas}} [q_0, \xi] = \int^{ \beta }_{0}
  d\tau  \, \delta (\tau ,\tau ) \frac{1}{6} R_{\mu \nu } (q_0)
  \xi^\mu(\tau )\xi^ \nu(\tau ) +\dots~.
\label{ins3.15}\end{eqnarray}

Collecting all terms, we obtain
the desired expansion of the
partition function (\ref{ins2.6})
in terms of the coordinates $\xi^\mu(\tau )$
around the origin
\begin{eqnarray}
 {\cZ}^{\rm P} =   \oint {\cal D}^D
\xi (\tau)\sqrt{g (q_0)}\,e^{-{\cal A}_{\rm e}^{(0)} [q_0, \xi]
 - {\cal A} _{{\rm e}}^{\rm int}[q_0, \xi]},
\label{ins3.16}\end{eqnarray}
with the total interaction
\begin{eqnarray}
{\cal A}_{{\rm e}}^{\rm int} [q_0, \xi] = {\cal A}_{\rm e}^{{\rm int},4}
[q_0, \xi] + {\cal A}_{\rm e}^{{\rm meas}} [q_0, \xi],
\label{ins3.17}\end{eqnarray}
and the measure written down in the notation (\ref{ins2.4}).
%where $\xi^\mu(\tau ) =0$ is the classical saddle-point and,

The path integral (\ref{ins3.16}) cannot immediately
be calculated perturbatively in the standard way, since
the quadratic form of the free action (\ref{ins3.6})
is degenerate. The spectrum of the operator $-d_\tau ^2$
in the space of periodic functions $\xi^\mu(\tau )$
has a zero mode. The zero mode is associated with the
fluctuations of the temporal average of $\xi^\mu (\tau )$:
\begin{equation}
\xi_0^\mu=\bar \xi^\mu \equiv
\hbeta^{-1}\int_0^{\hbeta} d \tau \,\xi^\mu(\tau ).
\label{ins@vta}\end{equation}
Small fluctuations of $\xi_0^\mu$ have the effect of moving
the path as a whole infinitesimally through the manifold.
The same movement can be achieved by
changing $q_0^\mu$ infinitesimally.
Thus we can replace the integral over the path average
$ \xi_0^\mu$ by an integral over $q_0^\mu$,
provided  that we properly account for the change of
measure arising from  such a variable transformation.

Anticipating such a change,
the path average (\ref{ins@vta})
can be set equal to zero eliminating the
zero mode in the fluctuation spectrum.
The basic free correlation function
$\langle \xi^\mu (\tau ) \xi^ \nu (\tau ')\rangle$
can then easily be found
from its spectral representation.
We solve the trivial eigenvalue problem of
the operator $- d_\tau^2 $
in the quadratic action (\ref{ins3.6}):
\begin{eqnarray}
  -d_\tau ^2\, u_m (\tau ) =  \lambda _m u_m (\tau ),~~~
\label{ins3.18}\end{eqnarray}
and impose periodic boundary conditions $u_m (0) = u_m ( \beta )$.
The obvious eigenfunctions are
%
%\begin{eqnarray}\!\!\!\!\!\!\!\!
$u_m (\tau ) =  e^{-i \omega_m \tau },$
%\quad\quad  \omega_m = \frac{2 \pi m}{ \beta },~~~~m
% = 0, \pm1,\,\pm2,\,\dots ~\,,
%\label{ins3.19}\end{eqnarray}
%
where $\omega _m = {2 \pi m}/{ \beta }$ are the Matsubara frequencies
with $m = 0,\,\pm 1,\,\pm2,\, \dots~$.
The eigenvalues are
$\lambda _m = \omega_m^2$.
The eigenfunctions are orthonormal,
\begin{eqnarray}
\frac{1}{\beta} \int_{0}^{\beta} d\tau\,
 u_{m}^* (\tau ) u_{m'} (\tau ) =  \delta _{m,m'}.
\label{ins3.20}\end{eqnarray}
and satisfy the completeness relation
\begin{eqnarray}
\frac{1}{\beta}\sum_{m} u_{m}^* (\tau ) u_m (\tau ')
 =  \delta (\tau - \tau ').
\label{ins3.21}\end{eqnarray}
Fixing $\xi^\mu_0=0$
removes the troublesome eigenmode $\lambda_0 = 0$
from the spectral representation.
This leads to the correlation function
\begin{equation}
\langle \xi^\mu (\tau ) \xi^ \nu (\tau ')\rangle^{q_0}
=g^{\mu \nu } (q_0)[-d_\tau ^2]^{-1} (\tau -\tau ')
=  g^{\mu \nu } (q_0) \Delta '(\tau ,\tau '),
\label{ins@propa0}\end{equation}
where $\Delta '(\tau ,\tau ')$ is the Green function
of the operator $- d_\tau ^2$ without the zero mode:
\begin{equation}
 \Delta '(\tau ,\tau ')
= \frac{1}{\beta} \sum _{m\neq0}\frac{u_m^* (\tau ) u_m (\tau ')}{\lambda_m}
  = \frac{1}{ \beta }\sum _{m\neq0}
\frac{e^{-i \omega _m(\tau -\tau ')}}{\omega _m^2}.
\label{ins@propa}\end{equation}
Performing the sum with the help of the formula
\begin{equation}
\sum_{m=1}^{\infty}\frac{\cos m \tau}{m^2}
  = \frac{1}{6}\pi^2 - \frac{1}{2}\pi|\tau|
+ \frac{1}{4}\tau^2  ,~~\tau \in [0, \beta ),
\end{equation}
yields for $\tau ,\tau '$ in the presently relevant interval
$[0, \beta )$ the translationally invariant expression:
\begin{equation}
 \Delta '(\tau ,\tau ') = \Delta '(\tau -\tau ')
\equiv \frac{|\tau -\tau '|^2}{2 \beta }
-\frac{|\tau -\tau '|^2 }{2} + \frac{\beta }{12}\,.~~~
\label{ins@propag}\end{equation}
which satisfies the inhomogeneous differential equation
\begin{eqnarray}
 - d_\tau ^2 \, \Delta ' (\tau -\tau ')
 =  \bardelta (\tau - \tau')\,,
\label{ins5.33}\end{eqnarray}
where
the right-hand side
contains
 an extra term on the right-hand side
due to the missing zero eigenmode in the
spectral representation:
\begin{eqnarray}
\bardelta  (\tau -\tau ') \equiv \frac{1}{\beta} \sum _{m\neq0}
e^{-i \omega _m(\tau -\tau ')}  = \delta (\tau -\tau ')
- \frac{1}{\beta }\,.
\label{ins5.33a}\end{eqnarray}
Both $ \delta '(\tau -\tau ')$ and
$ \Delta '(\tau -\tau ')$
are periodic
in the interval $\tau -\tau '\in [0,\hbar  \beta )$.
\section{Arbitrariness of $q_0^\mu$}
We now take advantage of an important property
of the perturbation expansion of the partition function (\ref{ins3.16})
around $q^\mu (\tau ) = q^\mu_0$:
the {\em independence\/} of the choice of $q_0^\mu$.
The separation (\ref{ins3.1}) into  a constant $q_0^\mu$ and
a time-dependent $\xi^\mu (\tau )$  paths
must lead to the same result
%expansion in powers of $\beta$
for any nearby constant $q_0'{}^\mu$ on the manifold.
The result must therefore be invariant under
an arbitrary infinitesimal displacement
\begin{eqnarray}
q^\mu_0\rightarrow q_{0 \varepsilon }^\mu = q^\mu_0 +  \varepsilon ^\mu,
\quad  | \varepsilon |\ll 1.
\label{ins4.1}\end{eqnarray}
In the path integral, this will be
compensated by some
translation of fluctuation  coordinates $\xi^\mu(\tau )$,
which will have the general nonlinear form
\begin{eqnarray}
\!\!\!\!\!\!\!\!\!\!
\xi^\mu \rightarrow \xi_ \varepsilon ^\mu = \xi^\mu
 - \varepsilon ^ \nu Q_ \nu ^\mu (q_0, \xi).~~
\label{ins4.2}\end{eqnarray}
The transformation matrix $Q_\nu ^\mu (q_0, \xi)$
satisfies the obvious initial condition
$Q_ \nu ^\mu (q_0, 0) =  \delta _ \nu ^\mu$.
The path  $q^\mu(\tau ) = q^\mu(q_0,\xi(\tau ))$
must remain invariant under simultaneous transformations
(\ref{ins4.1}) and (\ref{ins4.2}), which  implies that
\begin{eqnarray}
 \delta q^\mu \equiv
 q_ \varepsilon ^\mu - q^\mu =
 \varepsilon ^ \nu  D_ \nu  q^\mu (q_0, \xi) = 0\,,
\label{ins4.3}\end{eqnarray}
where $D_\mu$ is the infinitesimal transition operator
\begin{eqnarray}
  D_\mu = \frac{\partial}{\partial q_0 ^\mu}  - Q_\mu^ \nu
  (q_0, \xi) \frac{\partial}{\partial \xi^ \nu }\,.
\label{ins4.4}\end{eqnarray}
Geometrically, the matrix $Q_\nu^ \mu (q_0,\xi)$ plays the role
of a locally flat nonlinear connection \cite{ins9}.
It can be calculated as follows.
We express the vector $q^\mu (q_0, \xi)$
in terms of the geodesic coordinates $\xi^\mu$ using
Eqs.~(\ref{ins3.1}), (\ref{ins3.2}), and (\ref{ins3.3}), and substitute
this into
Eq.~(\ref{ins4.3}). The coefficients of
$ \varepsilon ^ \nu $ yield the equations
\begin{eqnarray}
   \delta ^\mu_ \nu  + \frac{\partial \eta^\mu (q_0, \xi)}
{\partial q_0^ \nu }  -  Q_ \nu ^ \kappa  (q_0,\xi)
  \frac{\partial \eta^\mu (q_0, \xi)}{\partial \xi^ \kappa }=0,
\label{ins4.5}\end{eqnarray}
where by Eq.~(\ref{ins3.2}):
\begin{eqnarray}
 \frac{\partial \eta^\mu (q_0,\xi)}{\partial q_0^ \nu } =
 -\frac{1}{2} \partial_ \nu   \Gamma ^\mu_{( \sigma \tau )}
 (q_0) \xi^ \sigma \xi^\tau  - \dots~,
\label{ins4.6}\end{eqnarray}
and
\begin{eqnarray}
 \!\!\!\!\!\!\!\!\!\!\!\!\frac{\partial \eta^\mu (q_0 , \xi)}
 {\partial \xi^ \nu } &=&
   \delta ^\mu_ \nu  -  \Gamma ^\mu_{( \nu  \sigma )} (q_0) \xi^ \sigma
  - \frac{1}{2}  \Gamma ^\mu_{ (\nu  \sigma \tau) } (q_0) \xi^ \sigma
  \xi^\tau - \dots \nonumber \\
& = & \delta _ \nu ^\mu -  \Gamma ^\mu_{ \nu  \sigma } \xi^ \sigma  -
\frac{1}{3} \left(\partial_ \sigma  \Gamma ^\mu_{ \nu \tau } + \frac{1}{2}
 \partial_\nu  \Gamma ^\mu_{ \sigma \tau }
 - 2 \Gamma ^ \kappa _{ \tau  \nu }
 \Gamma ^\mu_{ \kappa \sigma } -  \Gamma ^ \kappa _{\tau  \sigma }
   \Gamma ^\mu_{ \kappa  \nu }\right) \xi ^ \sigma \xi ^\tau - \dots\,.
\label{ins4.7}\end{eqnarray}
To find $ Q^\mu_ \nu (q_0, \xi)$, we invert the expansion (\ref{ins4.7}) to
\begin{eqnarray}
 \!\!\!\!\!\!\! \!\!\!\!
\left[ \left(\frac{\partial\eta (q_0, \xi)}
{\partial \xi}\right)^{-1}\right] ^\mu_{~ \nu}
\!\!\!\!\!\!\!& = & \!\!\!\delta _ \nu ^\mu +  \Gamma ^\mu_{ \nu  \sigma }
 \xi^ \sigma\!  +
\frac{1}{3} \left(\partial_ \sigma  \Gamma ^\mu_{ \nu \tau } + \frac{1}{2}
 \partial_\nu  \Gamma ^\mu_{ \sigma \tau }
\! + \Gamma ^ \kappa _{ \tau  \nu }
 \Gamma ^\mu_{ \kappa \sigma }\! -  \Gamma ^ \kappa _{\tau  \sigma }
\Gamma ^\mu_{ \kappa  \nu }\right) \xi ^ \sigma \xi ^\tau\! +\! \dots
\label{ins4.8}\\
\!\!\!&=&\!\!\!
 \left(\frac{\partial \xi^\mu (q_0, \eta)}{\partial \eta^ \nu }\right)_{
   \eta = \eta (q_0, \xi)}\,,
\nonumber\end{eqnarray}
the last equality indicating that the result (\ref{ins4.8})
can also be obtained from the inverted expansion (\ref{ins3.2}):
\begin{eqnarray}
\xi ^\mu (q_0, \eta) = \eta^\mu
 + \frac{1}{2}\tilde\Gamma_{(\sigma \tau) }{}^\mu
(q_0) \eta ^\sigma \eta^\tau
+ \frac{1}{6}\tilde \Gamma _{(\sigma \tau \kappa)}{}^\mu
   (q_0) \eta^ \sigma \eta^\tau  \eta^\kappa + \dots,
\label{ins4.9}\end{eqnarray}
with
\begin{eqnarray}\!\!\!\!\!\!\!\!\!\!
\tilde\Gamma_{\sigma \tau }{}^\mu (q_0)  &=&
\Gamma_{ \sigma \tau }{}^\mu ,\nonumber \\
\!\!\!\!\!\!\!\!\!\!
\tilde\Gamma _{ \sigma \tau  \kappa }{}^\mu (q_0)  &=&
\Gamma _{\sigma \tau \kappa}{}^\mu
 + 3\Gamma _{\kappa \sigma }{}^\nu \Gamma _{\nu \tau }{}^\mu
 %\nonumber \\ &&
 = \partial_ \kappa  \Gamma _{ \sigma \tau }{}^\mu
 +  \Gamma_{ \kappa  \sigma } {}^\nu
 \Gamma _{ \nu \tau }{}^\mu , \nonumber\\&\vdots& ~.
\label{ins4.9a}\end{eqnarray}
Indeed, differentiating (\ref{ins4.9}) with respect to $\eta^\nu$,
and reexpressing the result in terms of $\xi^\mu$ via Eq.~(\ref{ins3.2}),
we find once more (\ref{ins4.8}).

Multiplying both sides of Eq.~(\ref{ins4.5}) by
(\ref{ins4.8}), we express the nonlinear connection
$Q^\mu_ \nu (q_0,\xi)$  by means of geodesic
coordinates $\xi^\mu (\tau)$ as
\begin{eqnarray}
 Q^\mu_ \nu (q_0,\xi) =  \delta ^\mu_ \nu +  \Gamma ^\mu_{ \nu  \sigma }
(q_0) \xi^ \sigma + \frac{1}{3} R_{ \sigma  \nu \tau }{}^\mu
  (q_0) \xi^ \sigma  \xi^\tau  + \dots ~~.
\label{ins4.10}\end{eqnarray}
The effect of simultaneous transformations (\ref{ins4.1}), (\ref{ins4.2})
upon the fluctuation function
$\eta^\mu = \eta^\mu(q_0, \xi)$ in
Eq.~(\ref{ins3.2}) is
\begin{eqnarray}
\!\!\!\!\!\!\!\!\!\!
\eta^\mu \rightarrow \eta'{}^\mu = \eta^\mu -  \varepsilon ^ \nu
\bar Q_ \nu ^\mu (q_0, \eta),\quad \bar Q_ \nu ^\mu (q_0, 0)
 =  \delta ^\mu_ \nu ,
\label{ins4.11}\end{eqnarray}
where the matrix $\bar Q_ \nu ^\mu (q_0, \eta)$
is related to $Q_ \nu ^\mu (q_0, \xi)$ as follows
\begin{eqnarray}
 \bar Q_ \nu ^\mu (q_0, \eta) = \left[ Q^ \kappa _ \nu  (q_0, \xi)
  \frac{\partial \eta^\mu (q_0, \xi)}{\partial \xi^ \kappa }
   - \frac{\partial \eta^\mu (q_0, \xi)}{\partial q_0^ \nu }\right]_{
   \xi = \xi (q_0, \eta)}\,.
\label{ins4.12}\end{eqnarray}
Applying Eq.~(\ref{ins4.5}) to the right-hand side of Eq.~(\ref{ins4.12})
yields $\bar Q_ \nu ^\mu (q_0, \eta) = \delta ^\mu_ \nu $,
as it should to compensate the translation (\ref{ins4.1}).

\comment{At this point we observe that
since the fluctuation field $\xi^\mu(\tau )$
is parallel, its covariant derivatives vanish,
\begin{eqnarray}
  \nabla_ \nu  \xi^\mu = 0.
\label{ins4.15}\end{eqnarray}
This opens up a certain freedom in
fixing  the transformation matrix
$\bar Q_ \nu ^\mu (q_0, \eta) $.
It is always possible to add
the total covariant derivative of an arbitrary
function of $\xi^\mu(\tau )$, for instance
the covariant  derivative $\nabla_ \nu  \eta^\mu
(q_0, \xi)$. Thus we may choose
\begin{eqnarray}
\bar Q_ \nu ^\mu (q_0, \eta) =  \delta ^\mu_ \nu + \nabla_ \nu
\eta^\mu (q_0, \xi) \vert_{\xi = \xi (q_0, \eta)}\,,
\label{ins4.13}\end{eqnarray}
where
\begin{eqnarray}
\nabla_ \nu  \eta^\mu (q_0, \xi) =
\left[ \frac{\partial}{\partial q_0^ \nu }
 - \Gamma ^ \kappa _{ \sigma  \nu }  (q_0) \xi^ \sigma
\frac{\partial}{\partial \xi^ \kappa}\right] \eta^\mu +
 \Gamma ^\mu_{ \kappa  \nu } (q_0) \eta^ \kappa.
\label{ins4.14}\end{eqnarray}
Substituting Eq.~(\ref{ins3.2}) into (\ref{ins4.13}) and using
Eq.~(\ref{ins4.14}), we find a linear connection defined
up to covariant derivative
\begin{eqnarray}
 \bar Q^\mu_ \nu (q_0, \eta) =  \delta ^\mu_ \nu  - \frac{1}{2}
   t_{ \nu  \sigma \tau }{}^\mu (q_0) \xi^ \sigma \xi^\tau
  - \dots =  \delta ^\mu_ \nu  - \frac{1}{2} t_{ \nu  \sigma \tau }{}^\mu
  (q_0) \eta^ \sigma  \eta^\tau  - \dots~,
\label{ins4.16}\end{eqnarray}
where
\begin{eqnarray}
 t_{ \nu  \sigma \tau }{}^\mu (q_0) = \partial_ \nu
  \Gamma ^\mu_{ \sigma \tau }- 2  \Gamma ^ \kappa _{\tau  \nu }
     \Gamma ^\mu_{ \sigma  \kappa } +  \Gamma ^\mu_{ \kappa  \nu }
   \Gamma ^ \kappa _{\tau  \sigma }.
\label{ins4.17}\end{eqnarray}
}

The above independence of $q_0^\mu$
will be essential for  constructing the correct perturbation
expansion for the path integral (\ref{ins3.16}).
For some special cases of the Riemannian manifold, such as a
surface of sphere in $D+1$ dimensions
which forms a homogeneous space $O(D)/O(D-1)$, all points are
equivalent, and the local independence
becomes global. This will be discussed further in Section \ref{inssecQP}.
%%%
\section{Zero-Mode Properties}
\label{inssecZMP}
%%%%%
We are now prepared to eliminate the zero mode
by the condition of vanishing average $\bar \xi^\mu=0$.
As mentioned before, the vanishing fluctuation
$\xi^\mu(\tau ) = 0$ is obviously a classical saddle-point
for the path integral (\ref{ins3.16}). In addition, because of
the symmetry (\ref{ins4.2}) there exist other equivalent
extrema $\xi^\mu_\varepsilon (\tau ) = - \varepsilon ^\mu =$ const.
The $D$ components of $\varepsilon ^\mu$
correspond to $D$ zero modes which we shall eliminate
in favor of a change of $q_0^\mu$.
The proper way of doing this
is provided by the Faddeev-Popov procedure.
We insert into the path integral (\ref{ins3.16})
the trivial unit integral, rewritten with the help of (\ref{ins4.1}):
\begin{equation}
1=\int d^Dq_0 \,\delta ^{(D)}(q_{0 \varepsilon }-q_0)  =
\int d^Dq_0\,\delta ^{(D)}( \varepsilon ),
\label{ins@50}\end{equation}
and decompose the measure of path integration
over all periodic paths $\xi^\mu(\tau )$
into a product of an ordinary integral
over the temporal average $\xi_0^\mu=\bar\xi^\mu$, and a remainder
containing only nonzero Fourier components \cite{insPI5}:
\begin{eqnarray}
 \oint {\cal D}^D\xi(\tau )
=\int \frac{d^D\xi_0}{ \sqrt{2\pi \beta }^D }
\oint {\cal D}\hspace{1pt}'{\hspace{-1pt}}^D\xi(\tau )\,.
\label{ins@51}\end{eqnarray}
According to Eq.~(\ref{ins4.2}), the path average $\bar\xi^\mu$
is translated under $\varepsilon ^\mu$ as follows
\begin{equation}
\bar \xi^\mu\rightarrow
\bar \xi_{\varepsilon}^\mu=
\bar \xi{}^\mu- \varepsilon ^\nu\frac{1}{ \beta }
\int_0^\hbeta d\tau \,Q^\mu_ \nu (q_0,\xi(\tau )).
\label{ins@}\end{equation}
Thus we can replace
\begin{equation}
\int \frac{d^D\xi_0}{ \sqrt{2\pi \beta }^D }
\rightarrow
\int \frac{d^D \varepsilon }{ \sqrt{2\pi \beta }^D }\,\det \!\left[
\frac{1}{ \beta }\int_0^{\hbeta} d\tau\,Q^\mu_ \nu (q_0,\xi(\tau ))\right].
\label{ins@@52}\end{equation}
Performing this replacement in
(\ref{ins@51}) and performing the integral over $ \varepsilon ^\mu$
in the inserted unity (\ref{ins@50}),
we obtain the measure of path integration in terms of
$q_0^\mu$
and geodesic coordinates
of zero temporal average
\begin{eqnarray}
 \oint {\cal D}^D\xi(\tau )
&=&\int d^D q_0 \,
\oint \frac{d^D\xi_0}{ \sqrt{2\pi \beta }^D }
\delta ^{(D)} ( \varepsilon )
  \oint{\cal D}\hspace{1pt}'{\hspace{-1pt}}^D\xi(\tau )\nonumber \\
&=&
\int \frac{d^Dq_0}{ \sqrt{2\pi \beta }^D }
  \oint {\cal D}\hspace{1pt}'{\hspace{-1pt}}^D\xi(\tau )\,
\det \!\left[\frac{1}{ \beta }
\int_0^{\hbeta} d\tau\,Q^\mu_ \nu (q_0,\xi(\tau ))\right]
.
\label{ins@p52}\end{eqnarray}
The factor on the right-hand side is
the Faddeev-Popov determinant $ \Delta [q_0,\xi]$
for the change from $ \xi_0^\mu$ to $q_0^\mu$.
We shall write it as an exponential:
\begin{eqnarray}
 \Delta[q_0, \xi] = \det\left[\frac{1}{\beta}
 \int^ \beta _0 d\tau\,  Q_ \nu ^\mu (q_0, \xi) \right]
= e^{-{\cal A} _{\rm e}^{\rm FP} [q_0, \xi]} \, ,
\label{ins5.8}\end{eqnarray}
where ${\cal A}_{{\rm e}}^{\rm FP} [q_0, \xi]$ is an auxiliary action
accounting for the Faddeev-Popov determinant
\begin{eqnarray}
 {\cal A}_{{\rm e}}^{\rm FP} [q_0, \xi]\equiv
 - \rm tr \log \left[\frac{1}{\beta} \int^ \beta _0 d\tau\,  Q^\mu_ \nu
  (q_0, \xi)\right] \, ,
\label{ins5.9}\end{eqnarray}
which must be included into the interaction (\ref{ins3.17}).
Inserting (\ref{ins4.10}) into Eq.~(\ref{ins5.9}), we find
explictly
\begin{eqnarray}
 {\cal A}_{{\rm e}}^{\rm FP} [q_0, \xi] &=&
 - {\rm tr }\log \left[\delta ^\mu_ \nu  +
(3\beta) ^{-1}\int^ \beta _0 d\tau\,
  R_{ \sigma  \nu \tau }{}^\mu (q_0) \xi^ \sigma (\tau )\xi^\tau (\tau )+
 \dots \right] \nonumber\\
&=& \frac{1}{3 \beta } \int^ \beta _0 d\tau\,
    R_{\mu \nu } (q_0) \xi^\mu \xi^ \nu  + \dots ~\,.
\label{ins5.10}\end{eqnarray}
The contribution of this action will crucial for obtaining the correct
perturbation expansion of the path integral (\ref{ins3.16}).

\comment{
Note that $\bar \Delta (q_0, \eta)$ is defined by Eqs.~(\ref{ins5.7})
and (\ref{ins4.16}), (\ref{ins4.17}) on the surface (\ref{ins5.3}) as
\begin{eqnarray}
 \bar  \Delta (q_0, \eta)  = e^{\rm tr \log \left[  \beta  \int^ \beta _0
  d\tau\, \frac{1}{2} t_{ \nu  \sigma \tau } {}^\mu (q_0) \eta^ \sigma
   \eta^\tau  + \dots\right] }
 =  \beta ^D e^{-{\cal A}_{{\rm e}}^{\rm FP} [q_0, \eta]}
\label{ins5.11}\end{eqnarray}
thus yielding, instead of (\ref{ins5.10}), the noncovariant
contribution to the action
\begin{eqnarray}
  {\cal A}_{{\rm e}}^{\rm FP} [q_0, \eta] = \frac{1}{2 \beta } \int^ \beta _0
 d\tau\, \, t_{\mu \sigma \tau }{}^\mu (q_0) \eta^ \sigma \eta^\tau +
  \dots~,
\label{ins5.12}\end{eqnarray}
where
\begin{eqnarray}
t_{\mu \sigma \tau }{}^\mu (q_0) = \left(\partial_\mu
\Gamma ^\mu_{ \sigma \tau }
- 2  \Gamma ^\mu_{ \sigma  \kappa }  \Gamma ^ \kappa _{\mu\tau } +
  \Gamma ^\mu_{ \kappa \mu}  \Gamma ^ \kappa _{ \sigma \tau }\right).
\label{ins5.13}\end{eqnarray}
}

With the new interaction
\begin{eqnarray}
{\cal A} _{\rm e,new}^{\rm int} [q_0, \xi] =
  {\cal A}_{{\rm e}}^{\rm int} [q_0, \xi] + {\cal A}_{{\rm e}}^{\rm FP}
  [q_0, \xi]
\label{ins5.14}\end{eqnarray}
the partition function (\ref{ins3.16}) can be written as
a classical partition function
\begin{eqnarray}
  {\cZ}^{\rm P} =
\int\frac{ d^D q_0}{ \sqrt{2\pi \beta }^D }
   \sqrt{g(q_0)}\, e^{ -\beta V^{\rm eff\,cl}(q_0)}\,,
\label{ins5.15}\end{eqnarray}
where $V^{\rm eff\,cl}(q_0)$ is the curved-space version
of the effective classical partition function of Ref.~\cite{insx0}.
The effective classical Boltzmann factor
\begin{equation}
B(q_0)\equiv e^{ -\beta V^{\rm eff\,cl}(q_0)}
\label{ins@effclpf}\end{equation}
is given by the path integral
\begin{eqnarray}
%\,z^{\rm P} (q_0)
%e^{ -\beta V^{\rm eff\,cl}(q_0)}=
B(q_0)=
\oint{\cal D}\hspace{1pt}'{\hspace{-1pt}}^D\xi(\tau ) \sqrt{g(q_0)}
%\delta ^{(D)} \left[ \int^ \beta _0 d\tau\, \xi ^\mu(\tau ) \right]
  e^{- {\cal A} _{\rm e}^{(0)} [q_0, \xi]
- {\cal A}_{\rm e,new}^{\rm int}[q_0, \xi]}.
\label{ins5.16}\end{eqnarray}
Since the zero mode is absent in the fluctuations
on the right-hand side,
the perturbation expansion is now straightforward.
We expand the path integral (\ref{ins5.16})
in powers of the interaction (\ref{ins5.14}) around the
free Boltzmann factor
\begin{eqnarray}
%\tilde {\cZ}_0^{\rm P} (q_0) =
B_0 (q_0)
=
\oint{\cal D}\hspace{1pt}'{\hspace{-1pt}}^D\xi(\tau ) \sqrt{g(q_0)}
% \oint \left[ \prod_{\mu,\tau }
% d\xi^\mu (\tau ) \right]  \delta ^{(D)} \left[ \int^ \beta _0 d\tau\,
%  \xi^\mu (\tau ) \right]
 e^{- \int^ \beta _0 d\tau\,  \frac{1}{2} g_{\mu \nu } (q_0)
     \dot \xi^\mu \dot \xi^ \nu }
\label{ins5.17}\end{eqnarray}
as follows:
\begin{eqnarray}
B (q_0) = B_0 (q_0)
%\tilde {\cZ}^{\rm P} (q_0) =\tilde {\cZ}_0^{\rm P} (q_0)
 \left[ 1 -
  \langle {\cal A}_{\rm e,new}^{\rm int} [q_0, \xi]\rangle^{q_0}
  + \frac{1}{2} \langle {\cal A}_{\rm e,new}^{\rm int}
 [q_0, \xi] {}^2 \rangle^{q_0} - \dots \right] ,
\label{ins5.18}\end{eqnarray}
where the $q_0$-dependent correlation functions are defined by
the Gaussian path integrals
\begin{eqnarray}
 \langle \dots \rangle^{q_0} =
% \left[\tilde {\cZ}_0^{\rm P} (q_0)      \right] ^{-1}
B^{-1} (q_0)
   \oint {\cal D}\hspace{1pt}'{\hspace{-1pt}}^D \xi (\tau )
%\oint \left[\prod_{\mu,\tau } d\xi^\mu (\tau )\right]
%   \delta ^{(D)} \left[ \int^ \beta _0 d\tau\, \xi^\mu(\tau ) \right]
  \left[ \dots\right]^{q_0} e^{-{\cal A}_{\rm e}^{(0)} [q_0,\xi] } .
\label{ins5.19}\end{eqnarray}
By taking the logarithm of (\ref{ins5.17}), we obtain
directly a cumulant expansion for
the effective classical potential $V^{\rm eff\,cl}(q_0)$.

For a proper normalization of the
Gaussian path integral (\ref{ins5.17})
we diagonalize  the free action in the exponent
by changing the components of the fluctuations
\begin{equation}
\xi^\mu(\tau ) \rightarrow \xi^a (\tau ) = e^a{}_\mu(q_0)
\xi^\mu(\tau ),~ a = 1, \dots , D.
\label{ins@5.20}\end{equation}
The basis vectors $ e^a{}_\mu(q_0)$ are ``square-roots"
of the metric  at $q_0^\mu$:
\begin{eqnarray}
&& g_{\mu \nu } (q_0) = e^a{}_\mu (q_0) e^a{}_ \nu  (q_0), ~~~~~
   \sqrt{g(q_0)} = \det e^a{}_\mu (q_0) = \left[\det e_a{}^\mu
  (q_0)\right]^{-1},
\label{ins5.20}\end{eqnarray}
satisfying the orthogonality relation
$e^a{}{}_\mu (q_0)e^b{}^\mu (q_0)= \delta ^{ab}$.
Substituting $\xi^\mu(\tau ) = e_a{}^\mu (q_0) \xi^a (\tau )$
into Eq.~(\ref{ins5.17}) and taking into account that
\begin{eqnarray}
\oint {\cal D}\hspace{1pt}'{\hspace{-1pt}}^D\xi^\mu(\tau ) \sqrt{g(q_0)}=
\oint {\cal D}\hspace{1pt}'{\hspace{-1pt}}^D\xi^a(\tau ),
\label{ins5.21}\end{eqnarray}
we find
\begin{eqnarray}
%\tilde {\cZ}_0^{\rm P} (q_0) =
B _0(q_0) =
\oint{\cal D}\hspace{1pt}'{\hspace{-1pt}}^D\xi^a (\tau )
% \oint \left[ \prod_{i,\tau } d\xi^a (\tau )\right]
 %  \delta ^{(D)} \left[ \int^ \beta _0 d\tau\, \xi^a (\tau )\right]
 e^{-\int^ \beta _0  d\tau\, \frac{1}{2} (\dot \xi ^a)^2} .
\label{ins5.22}\end{eqnarray}
We have kept the superscript
in the measure of integration
to clarify which components of $\xi$  are being considered.
If we expand the fluctuations $\xi^a(\tau )$ into the eigenfunctions
$e^{-i \omega _m\tau }$  of the operator $- d_\tau ^2$
for periodic boundary conditions $\xi^a (0) = \xi^a ( \beta )$,
\begin{eqnarray}
\xi^a (\tau ) = \sum_{m} \xi^a_m u_m (\tau ) =
  \xi^a_0 + \sum_{m \neq 0} \xi^a_m u_m ( \tau ),~~
 \xi^a_{-m} = \xi^a_m{\!\hspace{1pt}\!}^*,~~m > 0,
\label{ins5.23}\end{eqnarray}
and substitute this into the path integral
(\ref{ins5.22}), the exponent  becomes
\begin{eqnarray}
 && - \frac{1}{2} \int^ \beta _0 d\tau\,  \left[\xi^a (\tau )\right]^2
  = -\frac{\beta}{2} \sum_{m \neq 0}  \omega ^2_m \,\xi_{-m} \xi_m
  = - \beta \sum_{m >0}   \omega _m^2\,\xi^ \alpha {}^*_m \xi^ \alpha _m\,.
\label{ins5.29}\end{eqnarray}
The measure has the Fourier decomposition \cite{insPI5}
\begin{eqnarray}
\oint{\cal D}\hspace{1pt}'{\hspace{-1pt}}^D\xi^a(\tau )=
\prod_{a,m>0} \int \frac{d \xi_m^{\rm re} d \xi_m^{\rm im}}{N_m}\,,
\label{ins5.24}\end{eqnarray}
where the normalization
factor $N_m$ regularize the divergent
product of eigenvalues $\lambda_m = \omega^2_m$.
The proper values are $N_m = \pi/\beta \omega _m^2$,
and we find after performing
the Gaussian integrals in (\ref{ins5.22})
the correct result
for a free-particle  Boltzmann factor
\begin{eqnarray}
%\tilde {\cZ}_0^{\rm P}
 B_0(q_0) = 1,
\label{ins5.27}\end{eqnarray}
corresponding to a vanishing effective classical potential
in Eq.~(\ref{ins@effclpf}).
As a consequence, the partition function
(\ref{ins5.15}) becomes
\begin{eqnarray}
  {\cZ}^{\rm P}=  \int\frac{ d^D q_0}{
\sqrt{2 \pi  \beta } ^{D}}
  \sqrt{g(q_0)} \, \,B(q_0),
\label{ins5.30}\end{eqnarray}
with the perturbation expansion

\begin{eqnarray}
B(q_0)
= 1
 - \langle {\cal A}_{{\rm e,new}}^{\rm int}
 [q_0, \xi] \rangle^{q_0}
 + \frac{1}{2} \langle {\cal A}_{{\rm e,new}}^{\rm int}
 [q_0, \xi] {}^2 \rangle^{q_0} - \dots ~.
\label{ins5.30p}\end{eqnarray}
The expectation values
on the right-hand side are to be
calculated with the help of Wick contractions
involving the basic correlation functions
of  $\xi^a(\tau )$
associated with the unperturbed action  in (\ref{ins5.22}):
\begin{eqnarray}
\langle \xi^a (\tau ) \xi^b  (\tau ') \rangle^{q_0}
= \delta ^{ab} \Delta '(\tau ,\tau ')\,,
\label{ins@basicpr}\end{eqnarray}
which are of course consistent with
(\ref{ins@propa0})
via Eqs.~(\ref{ins@5.20}) and (\ref{ins5.20}).
\section{Covariant perturbation expansion}
\label{insWick}
We now perform all possible Wick contractions of the
fluctuations $\xi^\mu (\tau )$ in the expectation values (\ref{ins5.30p})
using the correlation function (\ref{ins@propa0}).
We restrict our attention to the lowest-order terms
only, since all problems of previous treatments arise already there.
Making use of Eqs.~(\ref{ins@propag}) and (\ref{ins5.33}),
we find for the interaction (\ref{ins3.17}):
\begin{eqnarray}
\langle {\cal A}_{{\rm e}}^{\rm int} [q_0,\xi] \rangle^{q_0} &=&
\int^{ \beta }_{0}
 d\tau\,  \frac{1}{6} \left[ R_{\mu \lambda _1 \nu  \lambda _2} (q_0)
\langle \xi^ {\lambda_1} \xi^{ \lambda _2} \dot \xi^\mu \dot \xi^ \nu
\rangle^{q_0} +  \delta (\tau ,\tau ) R_{\mu \nu } (q_0) \langle \xi^\mu
\xi^ \nu  \rangle^{q_0} \right]\nonumber \\
&  = & \frac{1}{72}  R (q_0)  \beta ,
\label{ins5.36}\end{eqnarray}
and for (\ref{ins5.10}):
\begin{eqnarray}
\langle {\cal A}_{{\rm e}}^{\rm FP} [q_0, \xi] \rangle^{q_0}
 = \int^ \beta _{0}
 d\tau\,  \frac{1}{3 \beta } R_{\mu \nu } (q_0) \langle \xi^\mu \xi^ \nu
\rangle^{q_0} = \frac{1}{36} R(q_0)  \beta .
\label{ins5.37}\end{eqnarray}
The sum of the two contributions
yields the manifestly covariant
high-temperature expansion
up to two loops:
\begin{eqnarray}
B(q_0) = 1 - \langle {\cal A}_{{\rm e,new}}^{\rm int} [q_0, \xi]
   \rangle^{q_0} + \dots = 1 - \frac{1}{24} R (q_0)  \beta  + \dots
\label{ins5.35}\end{eqnarray}
in agreement with the partition function
density (\ref{ins2.5})
calculated from Dirichlet boundary conditions.
The associated partition function (\ref{ins5.30})
coincides with the partition function (\ref{ins2.1}).
Note the crucial role of the action (\ref{ins5.10})
coming from the Faddeev-Popov determinant
in obtaining the correct two-loop coefficient in Eq.~(\ref{ins5.35})
and the normalization in Eq.~(\ref{ins5.30}).

The intermediate transformation
to the geodesic coordinates $\xi^\mu (\tau)$
has made our calculations
rather lengthy if the action is given in arbitrary coordinates,
but it guarantees complete independence
of the coordinates in the result
(\ref{ins5.35}).
The entire derivation simplify, of course,
drastically if we choose from the outset
geodesic coordinates to parametrize the
curved space.
\section{Covariant result from noncovariant  expansion}
Having found the proper way of calculating
the  Boltzmann factor  $B(q_0)$
%It To compare our covariant result with earlier
%noncovariant treatments of Ref.~\cite{ins5}
we can easily set up a procedure for calculating
the same covariant result without the use of the geodesic
fluctuations $\xi^\mu(\tau )$.
Thus we would like to evaluate
the path integral (\ref{ins5.30})
by a direct expansion of the action
in powers of the noncovariant fluctuations $ \eta  ^\mu(\tau )$
in Eq.~(\ref{ins3.1}).
In order to  make $q_0^\mu$ equal to the path average,
$\bar q(\tau )$, we now require
$ \eta ^\mu(\tau )$ to have a vanishing temporal average $ \eta _0^\mu=
\bar \eta ^\mu=0$.

The expansions of the action (\ref{ins2.3}) and the measure (\ref{ins2.4})
in powers of $\eta^\mu (\tau)$
were already given in Eqs.~(\ref{ins3.5}), (\ref{ins3.8a}), (\ref{ins3.12})
and (\ref{ins3.13a}). The free action reads now,
\begin{eqnarray}
 {\cal A}_{\rm e}^{(0)} [q_0 ,\eta] = \int^{ \beta }_{0} d\tau  \frac{1}{2}
\eta^\mu(\tau ) [-g_{\mu \nu } (q_0)d_\tau^2 ] \eta^ \nu(\tau )\,,
\label{ins7.1}\end{eqnarray}
and the small-$\beta$ behavior of the path integral (\ref{ins5.30})
is governed by the interaction
\begin{eqnarray}
&&\!\!\!\!\!\!\!\!\!\!\!
\!\!\!\!
\!\!\!\!
\!\!\!\!
{\cal A}_{{\rm e}}^{\rm int} [q_0, \eta] = {\cal A}_{\rm e}^{{\rm int},4}
[q_0, \eta] + {\cal A}_{\rm e}^{{\rm meas}} [q_0, \eta]
\nonumber\\\!\!\!\!\!\!\!\!\!\!\!
& &\!\!\!\!\!\!\!\!= \frac{1}{2}\int_0^ \beta d\tau \left\{
\left[\partial_ \sigma g_{\mu \nu } (q_0) \eta^ \sigma
  + \frac{1}{2} \partial_ \sigma  \partial_\tau
 g_{\mu \nu } (q_0) \eta^ \sigma  \eta ^\tau
 \right] \dot \eta^\mu \dot \eta^\nu  - \delta (\tau, \tau)
 \partial_\sigma \Gamma ^\mu_{\tau \mu} (q_0)
\eta^ \sigma \eta^\tau \right\}.
\label{ins7.2}\end{eqnarray}

The  measure of functional integration
over  $\eta$-fluctuations without zero mode $ \eta ^\mu_0=\bar  \eta ^\mu$
can be deduced from the proper measure of nonzero
$\xi$-fluctuations:
\begin{eqnarray}
 \oint{\cal D}\hspace{1pt}'{\hspace{-1pt}}^D \xi (\tau )
 J (q_0, \xi) \Delta [q_0, \xi]
\equiv \oint {\cal D}^D\xi(\tau ) J (q_0, \xi)
\delta^{(D)} (\xi_0) \Delta [q_0, \xi].
\label{ins7.3}\end{eqnarray}
This is transformed to coordinates $\eta ^\mu (\tau)$ via Eqs.~(\ref{ins4.9})
and (\ref{ins4.9a})
yielding
\begin{eqnarray}
 \oint{\cal D}\hspace{1pt}'{\hspace{-1pt}}^D \xi (\tau )
 J (q_0, \xi) \Delta [q_0, \xi]
 = \oint{\cal D}\hspace{1pt}'{\hspace{-1pt}}^D \eta (\tau )
\bar \Delta [q_0, \eta]\,,
\label{ins7.4}\end{eqnarray}
where $\bar \Delta [q_0, \eta]$ is obtained from
the Faddeev-Popov determinant
$ \Delta [q_0, \xi]$ in Eq.~(\ref{ins5.8})
by expressed the coordinates $\eta^\mu (\tau)$
in terms of $\xi^\mu(\tau )$ and multiplying
the result with
a Jacobian  accounting for the change of
the $\delta$-function of $\xi_0$ to
a $\delta$-function of
$\eta_0$ via the transformation Eq.~(\ref{ins4.9}):
\begin{eqnarray}
\bar \Delta [q_0,\eta] &=&
\Delta [q_0, \xi (q_0, \eta)]
\times \det
\left(\frac{\partial \bar \eta^\mu (q_0, \xi)}{\partial \xi^ \nu }\right)_{
\xi = \xi (q_0, \eta)}.
\label{ins7.5}\end{eqnarray}
The last determinant has the exponential form
\begin{eqnarray}
\det
\left(\frac{\partial \bar \eta^\mu (q_0, \xi)}{\partial \xi^ \nu }\right)_{
\xi = \xi (q_0, \eta)} =
\exp\left\{ {\rm tr}\hspace{-1pt}\log\left[\frac{1}{\beta}
\int^{ \beta }_{0} d\tau\,
\left(\frac{\partial \eta^\mu (q_0, \xi)}{\partial \xi^ \nu } \right)
_{\xi =\xi (q_0, \eta)} \right]\right\}\,,
\label{ins7.6}\end{eqnarray}
where the matrix in the exponent has small-$\eta$ expansion
\begin{eqnarray}
&&\!\!\!\!\!\!\!\!\!\!\!\!\!\!\!\!\!\!\!\!\!\!\!\!\!\!\!\!
\left(\frac{\partial \eta^\mu (q_0, \xi)}{\partial \xi^ \nu }\right)_{
\xi = \xi (q_0, \eta)}
= \quad \delta _ \nu ^\mu  - \Gamma ^\mu_{ \nu  \sigma } \eta^ \sigma
\nonumber\\
&&
\quad\quad\quad\quad\quad -
\frac{1}{3}\left(\partial_ \sigma  \Gamma ^\mu_{ \nu \tau } + \frac{1}{2}
 \partial_\nu  \Gamma ^\mu_{ \sigma \tau }
 - 2\Gamma ^ \kappa _{ \tau  \nu }
 \Gamma ^\mu_{ \kappa \sigma } +\frac{1}{2}\Gamma ^ \kappa _{\tau  \sigma }
\Gamma ^\mu_{ \kappa  \nu }\right) \eta ^ \sigma \eta ^\tau + \dots\,.
\label{ins7.7}\end{eqnarray}
The factor (\ref{ins7.5}) leads to a new contribution
to the interaction (\ref{ins7.2}), if we rewrite
it as
\begin{eqnarray}
 \bar\Delta [q_0, \eta]  =
  e^{-\bar {\cal A}^{\rm FP}_{{\rm e}} [q_0, \eta]}\,.
\label{ins7.8}\end{eqnarray}
Combining Eqs.~(\ref{ins5.8}) and (\ref{ins7.6}), we find
a new Faddeev-Popov type action
for $\eta ^\mu$-fluctuations at vanishing $\eta _0^\mu$:
\begin{eqnarray}
\bar {\cal A}^{\rm FP}_{{\rm e}} [q_0, \eta]
& = & {\cal A}_{{\rm e}}^{\rm FP} [q_0, \xi(q_0,\eta)] -
 {\rm tr}\hspace{-1pt}\log\left[\frac{1}{\beta}
\int^{ \beta }_{0} d\tau\,
\left(\frac{\partial \eta^\mu (q_0, \xi)}{\partial \xi^ \nu } \right)
_{\xi = \xi (q_0, \eta)} \right] \nonumber\\
& = & \frac{1}{2 \beta } \int^ \beta _0
 d\tau\, \, T_{\sigma \tau }  (q_0) \eta^ \sigma \eta^\tau + \dots~,
\label{ins7.9}\end{eqnarray}
where
\begin{eqnarray}
 T_{\sigma \tau }{} (q_0) = \left(\partial_\mu
\Gamma ^\mu_{ \sigma \tau }
- 2  \Gamma ^\mu_{ \sigma  \kappa }  \Gamma ^ \kappa _{\mu\tau } +
  \Gamma ^\mu_{ \kappa \mu}  \Gamma ^ \kappa _{ \sigma \tau }\right).
\label{ins7.10}\end{eqnarray}
The unperturbed correlation functions associated with
the action (\ref{ins7.1}) are:
\begin{eqnarray}
 \langle \eta^\mu (\tau ) \eta^ \nu (\tau ')\rangle^{q_0}
 = g^{\mu \nu } (q_0) \Delta ' (\tau ,\tau ') \,
\label{ins7.11}\end{eqnarray}
and the free Boltzmann factor
is the same as in Eq.~(\ref{ins5.27}).
The perturbation expansion of the
interacting Boltzmann factor is to be calculated
from an expansion like (\ref{ins5.30p}):
\begin{eqnarray}
B(q_0)=
 1 - \langle {\cal A}_{{\rm e,new}}^{{\rm int}} [q_0, \eta] \rangle^{q_0}
 + \frac{1}{2}\langle
\left( {\cal A}_{{\rm e,new}}^{{\rm int}}[q_0, \eta]\right) ^2\rangle^{q_0}
 - \dots ~.
\label{ins7.15}\end{eqnarray}
where the interaction is now
\begin{eqnarray}
{\cal A} _{\rm e,new}^{\rm int} [q_0, \eta] =
  {\cal A}_{{\rm e}}^{\rm int} [q_0, \eta]
 + \bar {\cal A}^{\rm FP}_{{\rm e}} [q_0, \eta]\,.
\label{ins7.12}\end{eqnarray}
The Wick contractions of $\eta^\mu (\tau)$
are more numerous and complicated that
those of
the manifestly covariant coordinates $\xi^\mu (\tau)$.
In particular, the divergences containing
powers of $\delta (\tau ,\tau ) =  \delta (0)$
no longer cancel order
by order, but different orders conspire to remove them
in the final result. Consider for example
the first term in (\ref{ins7.15}):
\begin{eqnarray}
- \langle {\cal A}_{{\rm e}}^{\rm int} [q_0, \eta] \rangle^{q_0} &=&
 \frac{ \beta }{24}
 g^{ \sigma \tau } \left(\partial_ \sigma   \Gamma_{\tau \mu}^\mu +
 g^{\mu \nu }\Gamma _{\tau \mu, \kappa }\Gamma_{\sigma \nu}^\kappa
+ \Gamma _{\tau  \nu }^\mu \Gamma _{ \sigma \mu}^ \nu \right)
\nonumber\\
& - & \frac{ \beta }{24} g^{ \sigma \tau }
  \left( \partial_\mu  \Gamma _{ \sigma \tau }^\mu \!- 2
  \Gamma _{ \sigma  \nu }^\mu  \Gamma ^ \nu _{\mu\tau }\! +
  \Gamma ^\mu_{\mu \kappa } \Gamma _{\sigma \tau}^ \kappa \right)
\nonumber\\
&-& \frac{ \beta ^2}{24} \delta (0) g^{ \sigma \tau }\left( g^{\mu \nu }
\Gamma _{\tau \mu, \kappa }  \Gamma _{ \sigma  \nu }^  \kappa
 + \Gamma_{\tau \mu}^\nu  \Gamma ^ \mu _{ \sigma  \nu }\right)\,,
\label{ins7.13}\end{eqnarray}
where the term in the second line is the contribution
from the Faddeev-Popov action (\ref{ins7.9}).
The last term
contains the divergent quantity $\delta (0)$.
This is canceled by the same expression in the
second-order contribution to (\ref{ins7.15}):
\begin{eqnarray}
 \!\!\!\!\!\!\!\!\!\!\!\!\!\!\!\frac{1}{2}
 \left( {\cal A}_{{\rm e}}^{{\rm int}}[q_0, \eta]\right) ^2\rangle^{q_0} =
&-& \frac{ \beta }{24}  g^{ \sigma \tau } \left( g^{\mu \nu }
\Gamma _{\tau \mu, \kappa }  \Gamma _{ \sigma  \nu }^ \kappa
 + 2 \Gamma _{\tau  \nu }^\mu  \Gamma _{ \sigma \mu}^ \nu \right)
\nonumber \\
&+& \frac{ \beta ^2}{24}  \delta (0) g^{ \sigma \tau }
 \left(g^{\mu \nu }  \Gamma _{\tau \mu, \kappa }
 \Gamma ^ \kappa _{ \sigma  \nu } + \Gamma_{\tau \mu}^\nu
\Gamma ^\mu_{ \sigma  \nu } \right) .
\label{ins7.14}\end{eqnarray}
In calculating this, there is an additional complication
caused by the appearance
of initially undetermined integrals over products of distributions
of the type
$\int d\tau\, \,\epsilon^2 (\tau ,\tau )  \delta (\tau ,\tau )$.
Such integrals are determined uniquely
by the new calculus of distributions
in one dimension developed
in Ref.~ \cite{ins3}
from the coordinate invariance of path integrals.

The sum of Eqs.~(\ref{ins7.13}) and (\ref{ins7.14}) is of course finite
leading to the same covariant perturbation expansion as
before in Eq.~(\ref{ins5.35}).
Neglecting the contribution of the action (\ref{ins7.9})
as done by other authors in
Ref.~\cite{ins5}
will produce in Eq.~(\ref{ins7.15}) an additional noncovariant term
$g^{ \sigma \tau } T_{\sigma \tau } (q_0)/24 $.
This may be rewritten
as a covariant divergence  of a nonvectorial quantity
\begin{eqnarray}
 g^{ \sigma \tau } T_{\sigma \tau } = \nabla_\mu
  V^\mu,~~ V^\mu(q_0) = g^{ \sigma \tau } (q_0)
    \Gamma ^\mu_{ \sigma \tau } (q_0).
\label{ins7.15b}\end{eqnarray}
As such it does not contribute to the
integral over $q^\mu_0$ in Eq.~(\ref{ins5.30}),
but it is nevertheless a wrong noncovariant result
for the Boltzmann factor  (\ref{ins5.35}).

The appearance of a noncovariant term
in a treatment where $q_0^\mu$ is the path average of
$q^\mu(\tau )$ is not surprising.
If the time dependence of a path shows an acceleration,
the average of a path is not an invariant concept even
for an infinitesimal time.
One may covariantly
impose the condition of a vanishing temporal
average only upon fluctuation coordinates which
have no acceleration. This is the case of geodesic coordinates
$\xi^a(\tau )$ since their equation of motion at $q_0^\mu$ is
$\ddot\xi^a(\tau )=0$.
\comment{
to see that the noncovariant part of Eq.~(\ref{ins5.41})
is exactly compensated
by the expectation value of the determinant (\ref{ins5.12}), (\ref{ins5.13}):
\begin{eqnarray}
 - \langle {\cal A}_{{\rm e}}^{\rm FP} [q_0, \eta] \rangle^{q_0} =
- \int^ \beta _{0}
 d\tau\, \frac{1}{2 \beta } t_{\mu \sigma \tau }{}^\mu (q_0)
  \langle \eta^ \sigma \eta^\tau \rangle^{q_0} = -\frac{ \beta }{24}
  g^{ \sigma \tau } t_{\mu \sigma \tau }{}^\mu (q_0)
\label{ins5.42}\end{eqnarray}
in calculating the total expectation value
$ \langle {\cal A}_{\rm e,new}^{\rm int} [q_0, \eta] \rangle^{q_0} $
with ${\cal A}_{\rm e,new}^{\rm int} = {\cal A}_{{\rm e}}^{\rm int}
+ {\cal A}_{{\rm e}}^{\rm FP}$.
Its cancellation
against the expectation value (\ref{ins5.42})
allows us to obtain, however,
the same {\em manifestly covariant local\/}
partition function (\ref{ins5.35}) in the calculation used
the noncovariant split $q^\mu(\tau ) = q_0^\mu + \eta^\mu (\tau )$.
}
\section{Quantum particle on unit sphere}
\label{inssecQP}
A special treatment exits for particle in homogeneous spaces.
As an example, consider a quantum particle moving on a unit
sphere in a flat $D+1$-dimensional space.
The partition function is defined
by Eq.~(\ref{ins2.6}) with the euclidean action (\ref{ins2.3})
and the invariant measure (\ref{ins2.4}), where the metric
and its determinant are
\begin{eqnarray}
 g_{\mu \nu } (q) =  \delta _{\mu \nu } + \frac{q_\mu q_ \nu }{1 -q^2},
~~~~~~~g (q) = \frac{1}{1 - q^2}.
\label{ins6.1}\end{eqnarray}
It is, of course, possible to calculate
the Boltzmann factor $B(q_0)$ with the procedure
of Section \ref{inssecZMP}. Instead of doing this, we shall, however,
exploit the homogeneity of the sphere.
The invariance under reparametrizations of general Riemannian space
becomes here an isometry of the metric~(\ref{ins6.1}).
Consequently, the Boltzmann factor $B(q_0)$
in Eq.~(\ref{ins5.30}) becomes {\em completely independent\/}
of the choice of $q^\mu_0$, and the integral over $q_0^\mu$ in (\ref{ins5.15})
yields simply the total surface of the sphere times the
Boltzmann factor $B(q_0)$.

The homogeneity of the space
allows us to treat
paths $q^\mu(\tau )$ themselves as small quantum
fluctuations around the origin  $q^\mu_0 = 0$, which
extremizes the path integral (\ref{ins2.6}).
The possibility of this expansion is due to the fact
that at $q^\mu (\tau) = 0$, the movement is free
of acceleration, this being similar
to the situation in geodesic coordinates.

As before we now take account of the fact that there are
other equivalent saddle-points due to isometries
of the metric (\ref{ins6.1}) on
the sphere (see, e.g., \cite{ins10}).
The infinitesimal translations
of a small vector $q^\mu$:
\begin{eqnarray}
q_{\varepsilon}^\mu = q^\mu
+ \varepsilon ^\mu  \sqrt{1-q^2} ,
~~ \varepsilon ^\mu = {\rm const}, ~\mu =1, \dots , D
\label{ins6.2}\end{eqnarray}
move the origin $q^{\mu}_{0} = 0$ into $\varepsilon^\mu$.
%The vectors $l_\varepsilon$ can be found from
%$D$ infinitesimal rotations
%of a small vector $q^\mu$ into directions
%orthogonal to it:
%
%\begin{eqnarray}
% l_\varepsilon ^\mu (q) = \varepsilon ^\mu  \sqrt{1-q^2} ,
%~~ \varepsilon  ^\mu = {\rm const}, ~\mu =1, \dots , D.
%\label{ins6.3}\end{eqnarray}
%
Due to rotational symmetry, these fluctuations
have a vanishing action. They may be eliminated
from the path integral (\ref{ins2.6})
by including a
factor  $
\delta ^{(D)}(\bar q)$ to enforce a vanishing path average.
The associated Faddeev-Popov determinant $\Delta (q)$
is determined by the integral
\begin{eqnarray}
 \Delta (q) \int d^D  \varepsilon \,\delta ^{(D)}
 \left(\bar q_\varepsilon\right)
 = \Delta (q) \int d^D  \varepsilon \,\delta ^{(D)}
 \left(\varepsilon ^\mu  \frac{1}\beta\int^ \beta _0 d\tau\,
  \sqrt{1-q^2}\right) = 1\,.
\label{ins6.6}\end{eqnarray}
The result has the exponential form
\begin{eqnarray}
\Delta (q) = \left( \frac{1}\beta\int^ \beta _0 d\tau\,
 \sqrt{1-q^2}\right)^D = e^{-{\cal A}_{{\rm e}}^{\rm FP} [q]}\,,
\label{ins6.7}\end{eqnarray}
where ${\cal A}_{{\rm e}}^{\rm FP} [q]$
must be added to the action (\ref{ins2.3}):
\begin{eqnarray}
{\cal A}_{{\rm e}}^{\rm FP} [q]
 = - D \log \left(\frac{1}{ \beta }
 \int_0^ \beta  d\tau\,\sqrt{1 - q^2}\right).
\label{ins6.8}\end{eqnarray}
The Boltzmann factor $B(q_0)\equiv B$
is then given by the path integral without zero modes
\begin{eqnarray}
B = \oint\prod_{\mu,\tau }\left[ dq^\mu (\tau )
 \sqrt{g(q(\tau ))} \right] \delta ^{(D)} (\bar q)
 \Delta (q) e^{-{\cal A}_{\rm e} [q]} =
\oint{\cal D}\hspace{1pt}'{\hspace{-1pt}}^D q(\tau ) \sqrt{g(q(\tau ))}
  \Delta (q) e^{-{\cal A}_{\rm e} [q]}\,,
\label{ins6.5}\end{eqnarray}
where the measure
${\cal D}\hspace{1pt}'{\hspace{-1pt}}^D q$ is defined as
in Eq.~(\ref{ins5.24}). This can also be written as
\begin{eqnarray}
B = \oint {\cal D}\hspace{1pt}'{\hspace{-1pt}}^D q (\tau)
%\left(\prod_{\mu,\tau }
%dq^\mu (\tau ) \right) \delta ^{(D)} \left(\int^ \beta _0 d\tau\, q^\mu\right)
e^{-{\cal A}_{\rm e}[q] -{\cal A}_{{\rm e}}^J [q]-{\cal A}_{\rm e}^{\rm FP}
   [q] }\,,
\label{ins6.11}\end{eqnarray}
where ${\cal A}_{{\rm e}}^J [q]$ is a contribution
to the action (\ref{ins2.3}) coming from the product
\begin{eqnarray}
 \prod_\tau  \sqrt{g(q(\tau ))} \equiv  e^{-{\cal A} _{{\rm e}}^J [q]}.
\label{ins6.9}\end{eqnarray}
By inserting (\ref{ins6.1}), this becomes
\begin{eqnarray}
{\cal A}_{{\rm e}}^J [q] =
 - \int^ \beta _0 d\tau\, \frac{1}{2}  \delta (\tau ,\tau ) \log g (q) =
 \int^ \beta _0 d\tau\, \frac{1}{2}  \delta (\tau ,\tau ) \log (1- q^2)\,.
\label{ins6.10}\end{eqnarray}
The total partition function
is, of course,  obtained from $B$
by multiplication it with the
surface of the unit sphere in $D+1$ dimensions
$2\pi^{(D+1)/2}/\Gamma (D+1)/2$.

To calculate $B$ from (\ref{ins6.11}), we now expand
${\cal A}_{\rm e}[q], \,{\cal A}_{{\rm e}}^J [q]$
and ${\cal A}_{\rm e}^{\rm FP}[q]$ in powers of $q^\mu ( \tau )$.
The metric $g_{\mu \nu }(q)$ and its determinant $g(q)$ in Eq.~(\ref{ins6.1})
have the expansions
\begin{eqnarray}
 g_{\mu \nu } (q) =  \delta _{\mu \nu } + q_\mu q_ \nu  + \dots ~,~~~
  g(q) = 1 + q^2 + \dots~,
\label{ins6.12}\end{eqnarray}
and the unperturbed action reads
\begin{eqnarray}
  {\cal A}_{{\rm e}}^{(0)} [q] = \int^ \beta _0 d\tau\, \frac{1}{2}
\dot q^2 (\tau ).
\label{ins6.13}\end{eqnarray}
In the absence of the zero eigenmodes due to
the $\delta$-function over $\bar q$ in Eq.~(\ref{ins6.5}),
we find as in  Eq.~(\ref{ins5.27})
the free Boltzmann factor
\begin{eqnarray}
B_0 = 1.
\label{ins6.14}\end{eqnarray}
The free correlation function looks similar to (\ref{ins@basicpr}):
\begin{eqnarray}
 \langle q^\mu (\tau ) q^ \nu (\tau ')\rangle = \delta ^{\mu \nu }
  \Delta ' (\tau ,\tau ') .
\label{ins6.15}\end{eqnarray}
The interactions
coming from the higher expansions terms
in Eq.~(\ref{ins6.12}) begin with
\begin{eqnarray}
  {\cal A}_{{\rm e}}^{\rm int} [q] = {\cal A}_{\rm e}^{{\rm int},4} [q] + {\cal A}_{{\rm e}}^J[q] =
   \int^ \beta _0 d\tau\, \frac{1}{2} \left[ (q \dot q)^2 -  \delta
  (\tau ,\tau ) q^2 \right] .
\label{ins6.16}\end{eqnarray}
To the same order, the Faddeev-Popov interaction
(\ref{ins6.8}) contributes
\begin{eqnarray}
  {\cal A}_{{\rm e}}^{\rm FP} [q] = \frac{D}{2 \beta } \int^ \beta _0 d\tau\,  q^2.
\label{ins6.17}\end{eqnarray}
This has an important effect upon the two-loop perturbation expansion
of the Boltzmann factor
\begin{eqnarray}
B(q_0) = 1 -
\langle {\cal A}_{{\rm e}}^{\rm int} [q] \rangle^{q_0}
 - \langle {\cal A}_{{\rm e}}^{\rm FP} [q]\rangle^{q_0} + \dots~
 \equiv B.
\label{ins6.18}\end{eqnarray}
Performing the Wick contractions with the help of Eq.~(\ref{ins6.15}),
we find from Eqs.~(\ref{ins6.16}), (\ref{ins6.17}):
\begin{eqnarray}
\langle {\cal A}_{{\rm e}}^{\rm int} [q] \rangle^{q_0} =
 - \frac{D}{24}\beta
\label{ins6.19}\end{eqnarray}
and
\begin{eqnarray}
 \langle {\cal A}_{{\rm e}}^{\rm FP} [q]\rangle^{q_0} = \frac{D^2}{24}  \beta .
\label{ins6.20}\end{eqnarray}
Their combination in Eq.~(\ref{ins6.18}) yields
the high-temperature expansion
\begin{eqnarray}
 B = 1 - \frac{D(D-1)}{24}  \beta  + \dots~.
\label{ins6.21}\end{eqnarray}
This is in perfect agreement with Eqs.~(\ref{ins2.5}) and (\ref{ins5.35}),
since the scalar curvature for a unit sphere in $D+1$ dimensions is
$R = D(D-1)$. It is remarkable how
 the contribution (\ref{ins6.20}) of
the Faddeev-Popov determinant has made
the noncovariant result (\ref{ins6.19}) covariant.

%%%%%%%%%%%%%%%%%%%%%%%%%%%%%%

\end{document}